\documentclass[referee]{aa} 
\usepackage{graphicx,psfig}
\begin{document}
\title {\bf Are radio pulsars progenitors of AXPs and SGRs?
}
\titlerunning{Are radio PSRs progenitors of AXPs and SGRs?}
\author{F.K. Kasumov
\and
 A. O. Allahverdiev
\and
 A.I. Asvarov \inst{1}
}
\authorrunning{Kasumov et al.}

\offprints{F.K. Kasumov}

\institute {Institute of Physics
 of the National Academy of Sciences of Azerbaijan,\\ 33, H.Cavid Str, Baku AZ1143, Azerbaijan Republic\\
\email{astro@physic.ab.az}}

 \date{Received; accepted}

\abstract
{Possibility of realization of scenario for the AXPs and SGRs origins from radio pulsars, which were undergone frequent and strong glitches, was analyzed. It is shown that the characteristics of such pulsars - the possible progenitors of AXPs and SGRs - their association with supernova remnants (SNRs) and evolution on the $P - \dot P$ diagram, taking into account their real ages, conflict with offered scenario.

}

\maketitle

\section{Introduction}

Recently Lin \& Zhang (2004) offered the possible way of magnetars' origin according to
 which their progenitors may be standard radio pulsars (PSRs) exposed to frequent and 
 (once in several years) significant glitches (rapid spinups, sudden change of the period). 
Unlike the standard models, according to which there must be a significant 
difference in the pulsars' initial parameters, they assumed that PSR are born with 
a very close initial parameter  (period, magnetic field, est.) but exposed to different 
magnitude glitches. During the pulsar lifetime these glitches gradually  
lead to the growth of $P$ and $\dot P$ and therefore  (according to the standard formula 
$B\sim (P\dot P)^ {1/2}$) the value of the pulsars' magnetic field. Such scenario provides 
both the growth of   the pulsar's $P$ and $\dot P$ the small values of initial strength of the magnetic 
field $B \approx 3 \times 10^{10}  \div 3 \times 10^{11}$ $G$  to the values, 
characteristic for soft gamma ray repeaters (SGRs) and anomalous 
X-ray pulsars (AXPs) $B \approx 3 \times 13^{ 13}  \div 3 \times 10^{14}$ $G$, 
of a new class of NSs called magnetars. According to the estimation of Lin \& Zhang (2004) 
typical time, to reach the parameters of AXP and SGR is   $ \ge 2 \times 10^{ 5}$ $yr$ 
and the typical time for pulsars to enter the region of AXP and SGR on the diagram $B - P$ 
is equal   $ {\sim 1}{\rm .5} \times {\rm 10^{ 4} }  yr$. Therefore for the above mentioned scenario to be realized the following has to be:

1. Presence of permanent star quaking engine in the group of PSRs with very close physical parameters ($P, B, M,$ ets.) leading to sudden changes of $P$, the value and the frequency of which are considerably large $\Delta \dot P/\dot P \ge 0.0028$ and $\tau \sim  0.3$ per year.

2. Genetic relationship of magnetars and their possible   progenitors with SNRs, especially with those ages less than $10^5\,\,  \rm {yr}$, should not be found, because otherwise typical time of pulsars, with reasonable initial parameters and with the above mentioned values and frequencies of glitches, to enter the region of such objects is $t \ge 2 \times 10^5 \,\, yr$.

3.	The radio pulsars which are the possible progenitors of magnetars (AXPs and SGRs) in the way before the final reaching their region must show tendency of magnetic field increase with increase of period, i.e. must be pulsars with long periods  ($ P > 0.5\,\,sec$) and those with already raised magnetic fields ($B > 3 \times 13^{ 10} \,\,G$, to be disscused below).

4. On the $P - \dot P$ diagram the group of such pulsars must go to the uper corner of the diagram, i.e. they must show a positive correlation with each other as well as with age of pulsars.

According to date observational information we will analyze the answers to the question and demonstrate that there is no serious base for realization of such scenario for evolution of radio pulsars to AXPs and SGRs.

\section {Magnitude and frequency of glitches for possible progenitors of AXPs and SGRs.}

Modeling their scenario Lin and Zhang (2004) have used PSR J1757-24 with parameters of $ Ð = 0.25 ñåê$, $\dot P = 1.28 \times 10^{ - 13}$ and $B_0  = 2.6 \times 10^{11} \,\,G$. The possible genetic relations of this pulsar with SNR G 5.4-1.2 with age $\sim 10^{5}\,\, yr$ and the small speed of motion of this pulsar, the authors considered as an evidence of increase of the magnetic field in the course of its evolution.   
Indeed this pulsar showed a gigantic glitch  $\Delta \dot P/\dot P \ge 0.0037$  (Lane et al, 1996). However, it must be noted that in the number of works the discrepancy between the ages of pulsar and SNR is withdrawn either by the more precise estimation of SNR's age, taking into account the peculiarity of the environment in which it expanses (Gvamaradze, 2004), or by using the accretion fall-back disk model in estimation of pulsars$'$ ages (Marsden et al., 2002).
It is well known that many pulsars have glitches, especially the youngest ones such as Vela and Crab with the magnitudes differing considerably. To date for several hundred pulsars the glitches are detected, moreover, in 18 cases the glitches were rather large $
\Delta P/P \le 10^{ - 6} $, $\Delta \dot P/\dot P \le 10^{ - 5}  - 10^{ - 2}$  (Line et al., 2000) 

In our sample of pulsars - potential progenitors of magnetars, which includes $\sim 100$ objects, only in 4 cases the glitches have been detected. These are PSR J 1740-301 (with magnitude of the glitch of  $\Delta \dot P/\dot P = 0.0002 - 0.003$), PSR J 0528+2200 (0.00046), PSR J 1341-6220 (0.00015-0.003), PSR J 1801-2304  (0.00001). As we can see the magnitudes of  periods$'$ changes are comparable with the glitch of basic object PSR J 1757-24 in the model of Lin \& Zhang (2004), at that,  the frequency of glitches in these pulsars changes in the range of 1 to 0.2 per year. Therefore the parameters of glitches of the possible progenitors of AXPs and SGRs that we  included  in  our list rather close to the parameters in the Lin \& Zhang (2004) model.

\section {Association of possible progenitors of AXPs and SGRs with the SNRs.}
To date well known associations of AXPs 1E 2259+586, AX J1846-0258, 1E1841-045 and SNRs G109.1-1.0, G29.6-0.1, G274+0.0, respectively, are doubtless  (Gaensler 2004) Besides, according to Tagieva and Ankay (2003) the number of such possible associations can reach 6, but ages of SNRs in these pairs, with exception of the one (AXP 1Å 2259+58 with $t \approx 2 \times 10^5 \,\,yr$,and $\sim10^{3}-10^{4}\,\,yr$, respectively). Also, in the list of radio pulsars of possible progenitors of AXPs and SGRs (which must have $P \ge 0.5\,\,\sec$  è $B \ge 5 \times 10^{12} \,\,G$, there are 7 objects of genetically connected objects with hyper new remaining.  These are pairs PSR J1734-33 and G354.8-0.8, PSR J1119-61 and G229.2-0.5, PSR J1726-35 and G352.2-0.1, PSR J1632-48 and G336.1-0.2, PSR J1524-57 and G322.5-0.1, PSR J1124-59 and G229.0-1.8, PSR J1413-61 and G312.4-0.4 (Manchester et al., 2002). Age all of these SNRs do not exceed $10^{5}\,\,yr$. However, there is no significant difference between the estimation of pulsar's ages and ages of SNRs, which would be in favor of the offered model (as in case of basic pair PSR J1757-24 è G5.4 -1.2).

\section {Selection of radio pulsar - the potential progenitors of AXPs and SGRs, $B-P$ diagram.} 

As it was already pointed out, the small number of AXPs and SGRs (about 10 objects, in the Fig.1 and 2 they designed as ``+'') shows that even with the same birth rate of radio pulsars and AXPs and SGRs (which is limiting value for them) the number of the later is approximately as less as 1.5. Indeed, the relationship of radio pulsar's number $N_{PSR}  = R_{PSR}  \cdot t_{PSR}$   to the number of magnetars $N_M  = R_M  \cdot t_m $ at  $R_{PSR}  = R_M $ is proportional to $\sim t_{PSR} /t_M $, where $R$ and $t$ are birth rate and ages of these objects, respectively. Since $t_{PSR} /t_M  \approx 10^7 /10^5  = 100$ and observed ratio is$\sim  1400$ the number of AXPs and SGRs is by a factor of 1.5 less than the number of radio pulsars.  Taking into account the fraction of radio pulsars, "damaged" by glitches, the number of which among the known pulsar is  $\sim1/10$, we can conclude that the number of potential progenitors of AXPs and SGRs is less then all the observed pulsars by a factor of $\sim 15$, i.e. their amount should be about 100 objects.

On other hand, the canonical parameters of NSs and observed values of $P$ and $\dot P$ give for the values of the magnetic field $B = 3.2 \times 10^{19} (P\dot P)^{1/2}$ $\sim 10^{11}-10^{13}\, G$. If we also take into account the possible decay of the magnetic field with time $\tau _{\rm m} \sim 3\times 10^{6} \,yr$ (Guseinov et al. 2004) than their initial values can be thrice as much as.  For observed pulsars subjected to glitches to be potential progenitors of magnetars they must have the values of the magnetic field of $Â \sim (3 - 8)\times10^{12}\,\,G$. Finally, according to Ling \& Zhang (2004) at chosen parameters of glitch it is necessary $\sim 2\times 10^{5}\,yr$  for pulsar the full enter the AXP and SGR state and at the same the time to be in the region of AXPs and SGRs is  $ \sim 1.5\times 10^{4}\, yr$, i.e. objects subjected to initial increasing of magnetic field spent about $1/10$ time in before-magnetars stage. At initial value of $P_{0}\sim 10\, msec$ for this time the period of the pulsar can grow up to $0.5\, sec$

Taking into account the combined effect of all these factors we will restrict ourselves only to pulsars with $B \ge 5 \times 10^{12}\,\,G $ and  $P \ge 0.5\,\,\sec$, the number of which according to the catalogues of ATNF and Guseinov et al. (2002) equals $\sim90$.
The change of the magnetic field $B$ with period $P$ for current selection of objects is shown in the Fig.1.  For the mean value of pulsars velocity of $\sim 300\,\, km/s$ (Allahverdiev et al. 1997) the pulsars with age $< 10^{5}\,\,yr$ could move away from the birth place in the $ | Z |$ direction no far $100\,\, pc$, but during $\sim ~10^{6} yr$ their moving off could be   $ | Z | \ge 300\,\, pc$. That is why our sample consist of pulsars with $| Z | < 100 \,\,pc$  (relatively young pulsars) in Fig1 they designed by ``x'', but the old pulsars with   $| Z | > 300\,\, pc $ in the figure are shown as open circles ``o''. We have excluded from our analysis pulsars with $100\, pc < | Z |  < 300\, pc$  to make the difference in ages more prominent. To exclude possible selection effects and also to avoid inaccuracy in defining the distances to pulsars due to the possible deflection of pulsars$'$ birth places from the geometrical plane of the Galaxy (details in Allahverdiev et al.2005) relatively near pulsars ($ d<5\,\, kpc $) with $|Z|<100 \,pc$ and those with $|Z|>300\,\,pc$  are shown in Fig. 1 as ``squares'' and  ``black circles'', respectively.

\begin{figure}
\centering
\includegraphics[width=12 cm]{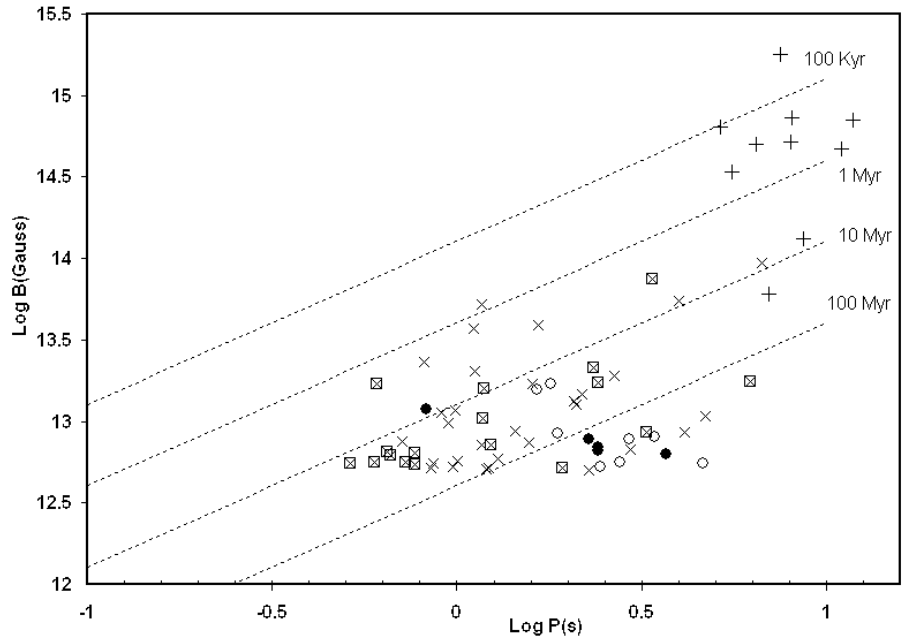}
\caption{$P-B$ diagram for  pulsars with $B>5\times 10^{12}\, G$ and $ P>0.5\, sec$. Straight lines are constant characteristic age $\tau$ lines. See text for details.}
\label{fig1}
\end{figure}

\begin{figure}
\centering
\includegraphics[width=12 cm]{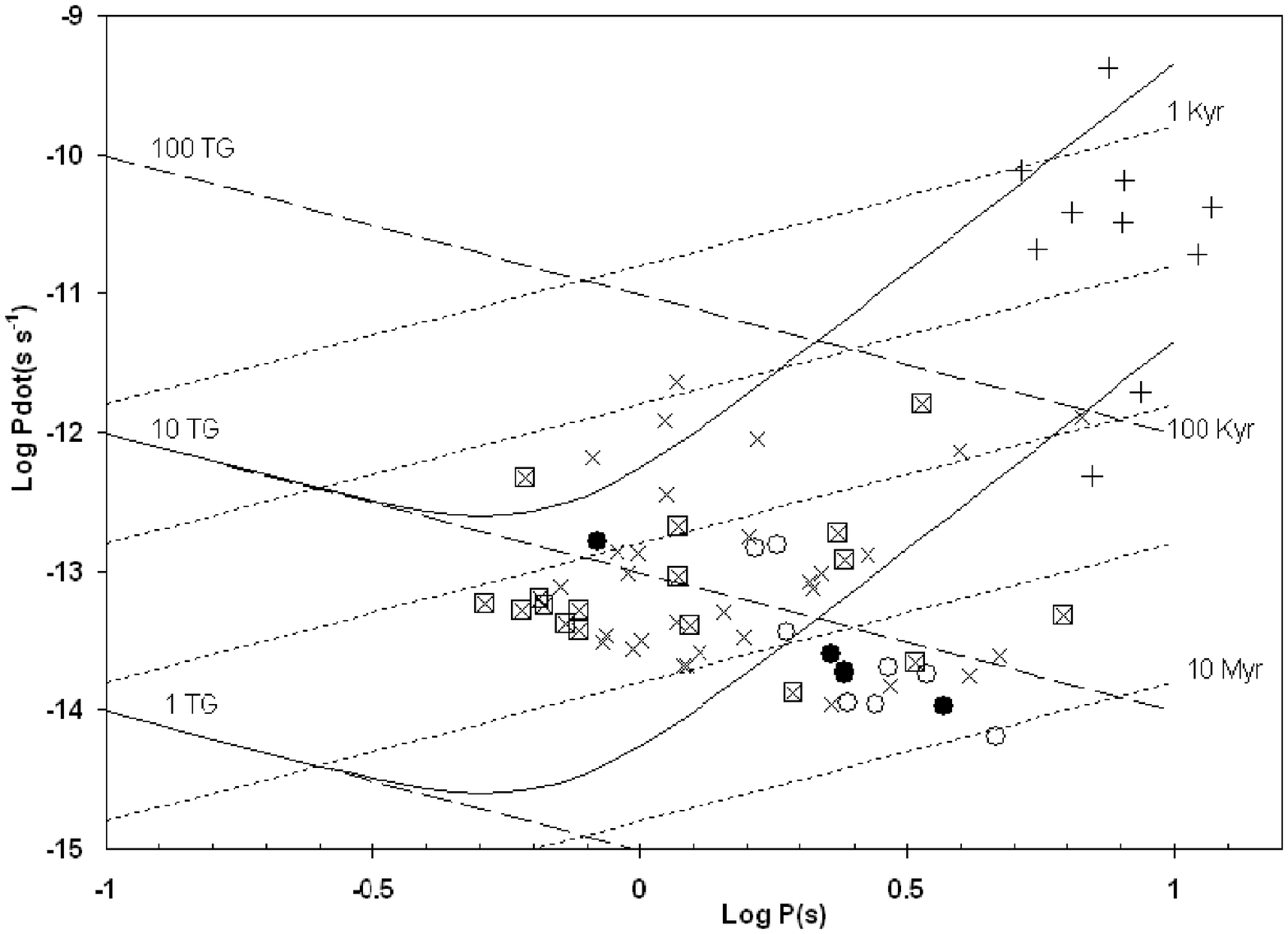}
\caption{ $P-\dot P$ diagram for pulsars with $B>5\times 10^{12}\, G$ and $ P>0.5\, sec$.  Solid lines correspond to lines of constant characteristic aget, dotted lines - constant magnetic field $B$. Solid-curved lines mark evolutionary track described by combine model.  Other symbols see in the text.  }
\label{fig2}
\end{figure}

\section {$P-\dot P$ diagram and possible evolutionary paths of progenitors of AXPs and SGRs} 

On the basis of our sample of radio pulsars - the possible progenitors of AXPs and SGRs (see Sec. IV) the $P-\dot P$ diagram is constructed (Fig.2). Symbols in this figure are the same as in Fig.1. Assume that these pulsars or part of them are actually exposed to pre-amplification of the magnetic field $B$ due-to glitches and achieved today values of $B$ and $P$ for $10^{4}\,\, yr$. In next $10^{5} -10^{6}\,\,yr$ years they must reach the position of AXPs and SGRs in the $P-\dot P$ diagram. In that case the distribution on the diagram should show the growth trend of their actual ages. Actual ages of pulsars are their kinematical ages, which should show the linear increase with distance from the Galactic plain  $(| Z |)$ taking into consideration the birth place of pulsars in different parts of the Galaxy and their deviation from the geometrical plane of the Galaxy (Hansen \& Pinney 1997; Berdnikov 1987).

As it can be seen in Fig.2 the growth trend of density of old objects with increasing $P$ doesn't observed both taking into consideration the selection effects and without them. Most likely we see an opposite picture: young pulsars $|Z| < 100\,\,pc$ nearly uniformly distributed up to value $P \ge 5\,\,sec$. Moreover, as is obvious from Fig.1 and 2 the growth of magnetic filed B is not observed with increasing of period P and actual pulsars age. Among pulsars with $|Z|>300\,\, pc$ there are only 3 objects with magnetic field more than $10^{13}\,\, G$ when the there is not limiting to distances and one such object when $d<5\, kpc$, whereas at $|Z|<100\,\, pc$ the number of pulsars with $B >10^{13}\,\,G$, and $P\geq 2\,\, sec$ equal 11 for the case without limiting on the distance and 4 taking into consideration $d < 5\,\,kpc$.

Of course, it is necessary to take into consideration the fact that most of the high-magnetic field radio pulsars are discovered in the low-altitude multibeam  Parks Galactic plane survey, and this circumstance can  definitely result in selection effect for high-latitude pulsars. Nevertheless, the data and information that we have to date does not confirm necessary evolutionary picture of the offered scenario about the origin of AXPs and SGRs.

Continuous curves in Fig. 2 describe pulsars$'$ evolutionary tracks in a combined ``dipole + propeller'' model offered by (Menou 2001) and Alpar et al. 2001 at various values of the initial magnetic field $B$ and the accretion speed. As can be seen, this model as in the case for all pulsars (Allahverdiev et al. 2005) describes neither the evolution process along propeller - dominant branch nor these highly -magnetized objects.

\section {Conclusions}

Thus, our study shows that any from above-mentioned general conditions (probably, excluding the first one) for realization of offered scenario is not confirmed by observational data. Observations give evidence in favour of standard, generally accepted, representation about the evolution of pulsars on $P-\dot P$ diagram  (Ruderman 2001).

Special non-standard picture of the evolution process of AXPs and SGRs continues to keep its own status (Tomson \& Duncan 1995; Malov et al. 2003). Thereupon it is important to note that the alternative propeller or fall-back disk model (Chatterjee et al. 2000; Alpar 2001) as well explain the origin of discrepancy between the characteristic and actual (kinematical) ages of pulsars (Marsden et al. 2001; Shu \& Hu, 2003). However and this offered model for an explanation of evolutionary tracks of all pulsars by combined action magnetic-dipole and propeller mechanisms are met with serious difficulties (Tagieva et al.; Allahverdiev et al. 2005). Probably in the light of discovery of X-rays from one of highly magnetized pulsar PSR J1718-37 $Â \sim 7.4\times 10^{13} \, G$ (Kaspi \& McLaughlin 2004), and as well as from the analysis of parameters of NSs and their possible time variation on the evolutionary tracks on the $P-\dot P$ diagram for pulsars, it is necessary to take into consideration others latent parameters of NSs (for example, masses (see, Kaspi \& McLaughlin 2004; Guseinov et al. 2005) in standard evolutionary scenarios.

\clearpage

\end{document}